\documentclass{cernyrep}
\usepackage[T1]{fontenc}
\usepackage[bookmarks, colorlinks=true, linktoc=page, pdftex, linkcolor=black, citecolor=black, urlcolor=blue]{hyperref}

\usepackage{fancyhdr}
\fancyhfoffset{4 mm}
\fancypagestyle{ARTTITLE}{%
\fancyhf{} 
\lhead{\hfill CERN Accelerator School Proceedings ---
{\it Medical Accelerators} ---  J\=urmala,  Latvia, 2026\hfill}
\lfoot{\hspace{3mm} Available online at \url{https://cas.web.cern.ch/previous-schools}}
\rfoot{\thepage\hspace*{3mm}}
 
}

\usepackage{varwidth}
\usepackage{xcolor}
\usepackage{wrapfig}
\usepackage{amsmath}

\begin{document}

\title{Special Applications of Superconducting Magnets}
\author{E. Felcini}
\institute{CNAO, Pavia, Italy}

\begin{abstract}

This paper explores the basic principles of superconducting magnet design for medical accelerators.
It briefly traces the history of superconductivity, discusses the critical parameters and their evolution over time, and introduces the concepts of operational margins, quench dynamics, and cooling strategies.
Next, it compares resistive and superconducting technologies and examines various magnet designs,
such as cos-theta, Canted Cos-Theta (CCT), superferric, and toroidal configurations. Finally, the
paper focuses on superconducting magnets for hadron therapy gantries, reviewing commercial
systems and cutting-edge initiatives, including the EuroSIG cos-theta, HITRIplus and i.FAST CCT, NIMMS superferric designs, and the GaToroid torus.
\end{abstract}

\maketitle

\thispagestyle{ARTTITLE}

\section{Introduction}

Superconducting magnets are, together with the accelerators, one of the key enabling technologies for modern particle therapy. Compared with resistive electromagnets, superconducting coils allow much higher magnetic fields to be generated in a given aperture, or equivalently, a given field to be generated with a much smaller and lighter magnet. This is of particular relevance for rotating gantries, whose size, weight, and cost scale directly with the beam rigidity and with the achievable bending field. This paper follows the structure of the lecture delivered at the CAS Topical Course on Medical Accelerators, and is organized into three parts: physical principles of superconductivity (Section~\ref{sec:supercond}), principles of superconducting magnet design (Section~\ref{sec:design}), and their application to gantries for particle therapy (Section~\ref{sec:gantries}). Concluding remarks are given in Section~\ref{sec:conclusions}.

\section{Principles of Superconductivity}
\label{sec:supercond}

\subsection{A brief history and the critical surface}

\begin{figure}[!htb]
	\centering
	\includegraphics[width=0.5\columnwidth]{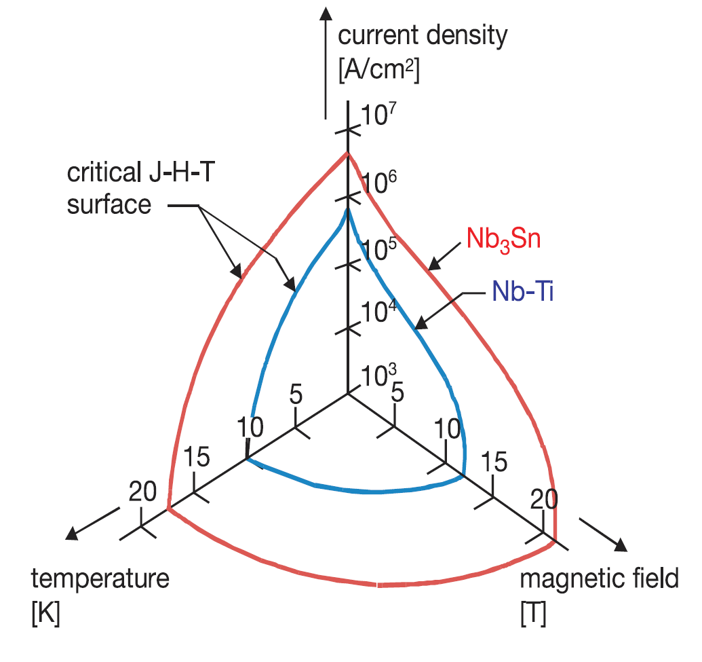}
        \caption{The critical $J$-$H$-$T$ surfaces of Nb-Ti and Nb$_3$Sn \cite{RossiTodesco2009}.}
	\label{fig:crit_surface}
\end{figure}

Superconductivity was discovered in 1911 by the Dutch physicist H. Kamerlingh Onnes, three years after he first liquefied helium (4.2~K) at the University of Leiden. Onnes observed that, below 4.2~K, the~electrical resistance of mercury became unmeasurably small, in fact zero rather than merely very small. A~couple of years later, at the Third International Congress of Refrigeration in Chicago in 1913, scientists were already speculating about 100,000~Gauss (10~T) magnets, cooled with a helium plant that would require only ``a relatively modest support''.  This optimistic prediction, however, was short-lived: the~superconducting state of mercury was found to disappear when an external field of only about 0.05~T was applied~\cite{Wilson2011}.
The reason for this apparent contradiction is that the superconducting state is bounded not by temperature alone but by a three-dimensional \emph{critical surface} in the space of temperature ($T$), current density ($J$) and magnetic field ($H$ or $B$). A material remains superconducting only if simultaneously
\begin{equation}
T < T_c \ \text{(critical temperature)}, \qquad J < J_c \ \text{(critical current density)}, \qquad B < B_c \ \text{(critical field)},
\end{equation}
and each superconducting material possesses its own critical surface, with operating regimes required to remain inside this surface. 
Figure~\ref{fig:crit_surface} shows the critical surfaces of Nb-Ti and Nb$_3$Sn, pointing out the~superior properties of the latter in terms of temperature, field, and current density. 

It took nearly a century of materials research and magnet engineering to develop a technology able to approach the field level of 10~T originally envisaged in the first years of superconductivity.

\subsection{Practical superconducting materials}

\begin{figure}[!htb]
	\centering
	\includegraphics[width=0.8\columnwidth]{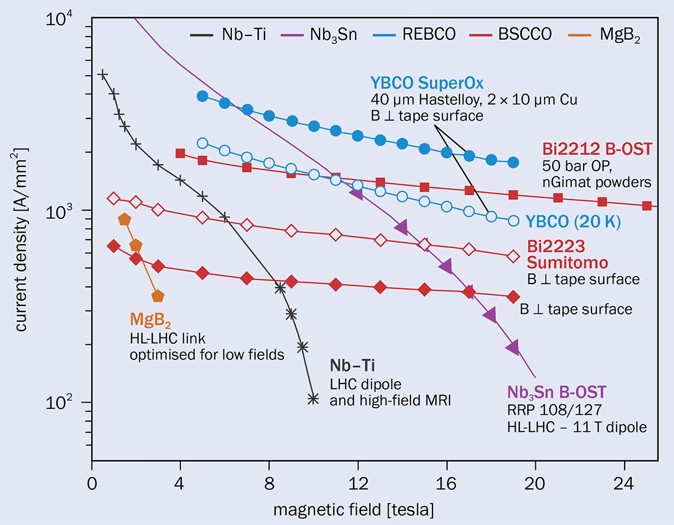}
        \caption{Engineering current density as a function of magnetic field for Nb-Ti, Nb$_3$Sn, MgB$_2$, YBCO/REBCO and BSCCO conductors at 4.2~K ~\cite{Todesco2026} (credit A. Ballarino).}
	\label{fig:J_B_4K}
\end{figure}

Following Onnes' discovery, the search for materials combining high critical temperature, current density and field led to the identification of a number of practical superconductors. Nb$_3$Sn (niobium-tin) was discovered in 1954, Nb-Ti (niobium-titanium) in 1962, and the high-temperature superconductor YBCO (yttrium barium copper oxide) in 1986~\cite{Ballarino2023}. Nb-Ti remains, to date, the workhorse conductor for accelerator magnets owing to its excellent mechanical ductility, which makes it comparatively easy to manufacture into multi-filamentary wires and cables.
As shown in Fig. \ref{fig:J_B_4K}, Nb$_3$Sn and the~high-temperature superconductors (HTS) REBCO/YBCO, and BSCCO extend the accessible field range beyond what Nb-Ti can provide, at the cost of increased engineering complexity and cost ~\cite{Ballarino2023,Todesco2026}.\\

The first Nb-Ti accelerator magnets ever built were eight quadrupoles with a peak field of 6.1~T, assembled at CERN in 1976 for an insertion of the Intersecting Storage Rings~\cite{Billan1976}. Large-scale industrialization of Nb-Ti magnet technology followed with the Fermilab R\&D programme, which led to the~700 dipoles (4.3~T, Nb-Ti) of the Tevatron, whose first beam was achieved in 1983~\cite{FNAL2026}. The LHC main dipoles, operating at 8.3~T and cooled to 1.9~K with superfluid helium, represent the current state of the art among Nb-Ti accelerator magnets; the first LHC main dipole prototype was built in 1993, and the full series of 1232 main dipoles was delivered in 2006~\cite{Ballarino2023}.
Nowadays, the CERN High-Luminosity and High Field Magnets programmes are developing Nb$_3$Sn and HTS magnets for LHC upgrades and for future colliders \cite{Todesco2025, Ambrosio_HiLumi, Bottura_HighField}.

\subsection{Operating margins}

Because the critical current density of a superconductor depends strongly on both temperature and field, the same conductor can be exploited very differently depending on the chosen operating point. For example, for the Nb-Ti conductor used in the LHC, an operating point at 5~K and 4~T supports a critical current density of the order of 1200~A/mm$^2$, whereas at 1.9~K and the same field of 4~T the critical current density rises to about 5000~A/mm$^2$~\cite{Bottura2013a}. The distance between the operating point and the critical surface, expressed either as a temperature margin or as a current margin, is therefore not a fixed conductor property but a design choice: the same conductor, operated at different temperatures, effectively yields two completely different magnets in terms of achievable current density and stability~\cite{Bottura2013a}.

\subsection{Quench and Stabilizer}

In the presence of a localized disturbance of thermal, mechanical, or electrical origin, a superconductor may approach or exceed its critical surface and then quench.
A quench is the transition of a superconducting magnet from its superconducting state into a normal, resistive state, associated with a sudden rise of voltage and temperature in the affected region of the winding~\cite{Bottura2013b}. Once quenched, a superconductor exhibits a comparatively high resistivity, and for this reason, it is convenient to introduce a~\emph{stabilizer}, usually copper, surrounding the superconducting material. When the operating current locally exceeds the critical current limit, the current is shared between the superconducting filaments and the surrounding stabilizer in the so-called current-sharing regime, characterized by a temperature lying between the~current-sharing temperature $T_{cs}$ and the critical temperature $T_c$~\cite{Bottura2013b}. If the heat generated in this process exceeds the heat that the cooling system is able to remove, the disturbance propagates, and the magnet quenches definitively; if instead the cooling is sufficiently effective and rapid, the magnet returns to stable superconducting operation~\cite{Bottura2013b}. A larger amount of copper stabilizer around the~superconducting filaments increases the stability margin of the magnet against such disturbances, but at the same time reduces its overall current density and therefore its efficiency \cite{Bottura2013b}.
Usually, for a finished cable, the critical current density is diluted by roughly 30\% due to the stabilizer, insulation, and other materials; this value is then referred to as the engineering current density, $J_\mathrm{eng}$~\cite{Todesco2026}.
The choice of the copper-to-superconductor ratio is a central trade-off in magnet design, together with the design of dedicated quench-detection and protection systems that are not covered in detail in this paper.

\subsection{Cooling options}

Superconducting magnets can be cooled using several technologies, differing in their thermal effectiveness and in their practicality for a given application~\cite{Gerbershagen2016}. Bath cooling in liquid helium at 4.2~K, or in superfluid helium at 1.9~K, provides the most effective cooling with excellent thermal stability, but requires the complex engineering of a rotating cryostat when applied to rotating structures such as gantries. Force-flow cooling, using supercritical helium at about 4.5~K and 3--8~bar, can be implemented either through dedicated channels machined in the magnet structure or directly inside cable-in-conduit conductors. Conduction cooling by cryocoolers transfers heat by conduction only, with a more limited heat-removal capability; this is nevertheless a typical choice for gantry magnets because of its practicality, even though crycooler efficiency may be reduced at certain rotation angles of the gantry. The choice of cooling scheme is therefore closely linked to the mechanical and operational constraints of the specific magnet design.

\section{Principles of Magnet Design}
\label{sec:design}
\subsection{Resistive versus superconducting magnets}
\begin{wrapfigure}{l}{0.4\textwidth}
\includegraphics[width=1\linewidth]{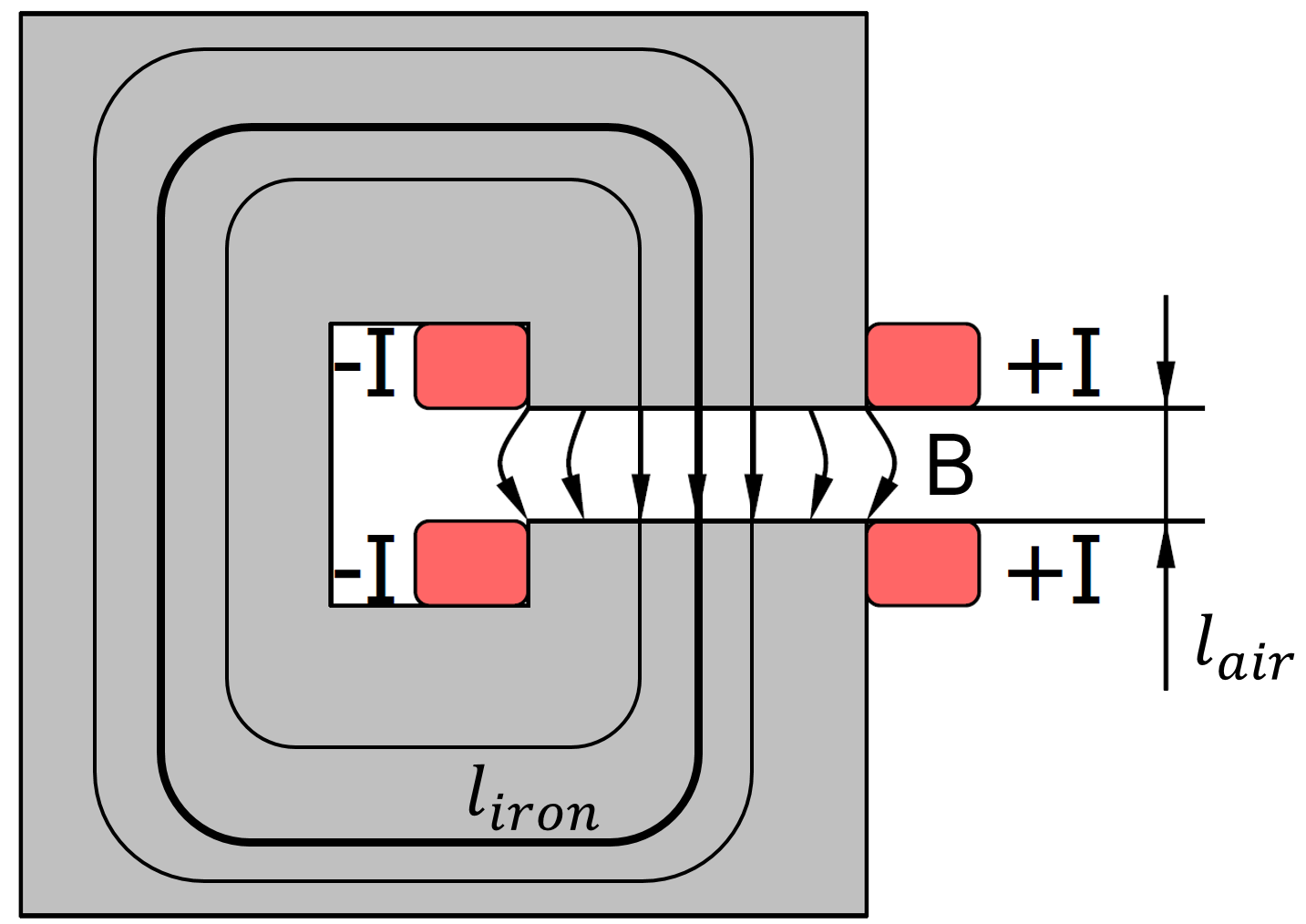} 
\caption{C-type resistive dipole ~\cite{BotturaMilaneseTommasini2020}.}
\label{fig:Resistive_Magnets}
\end{wrapfigure}

In a resistive iron-dominated magnet, the field in the aperture is set essentially by the shape of the iron yoke rather than by the coil geometry.
We can apply Ampere's law around a~closed loop that traverses the iron yoke and the air gap, as shown in Fig. \ref{fig:Resistive_Magnets}.

\begin{equation}
\begin{aligned}
 NI &= \oint H dl = H_{iron}l_{iron} + H_{air}l_{air} \\
&= \frac{B}{\mu_0\mu_r}l_{iron} + \frac{B}{\mu_0}l_{air}
\end{aligned}
\end{equation} \\ \\
Considering that the relative permeability of (unsaturated) iron is much greater than one ($\mu_r > 1000$, compared with $\mu_r = 1$ for air), one obtains 

\begin{equation}
B \approx \mu_0 \frac{N I}{l_{air}},
\end{equation}
where $l_{air}$ is the length of the air gap. \\

In a superconducting magnet, by contrast, the field is dominated by the shape of the coil rather than by any iron. The Biot--Savart law applied to a pair of infinitely long, oppositely energized conductors can be approximated as:
\begin{equation}
B \approx \mu_0 \frac{N I}{\pi r},
\end{equation}
where $r$ is the coil aperture radius. \\

A hybrid configuration, referred to as a \emph{superferric} magnet, combines a superconducting coil with an iron yoke, using the combination of coil and iron to shape the magnetic field; superferric magnets thus occupy an intermediate position between purely iron-dominated resistive magnets and purely coil-dominated superconducting magnets.

\subsection{Cos-theta magnets}

\begin{figure}[!htb]
	\centering
	\includegraphics[width=1\columnwidth]{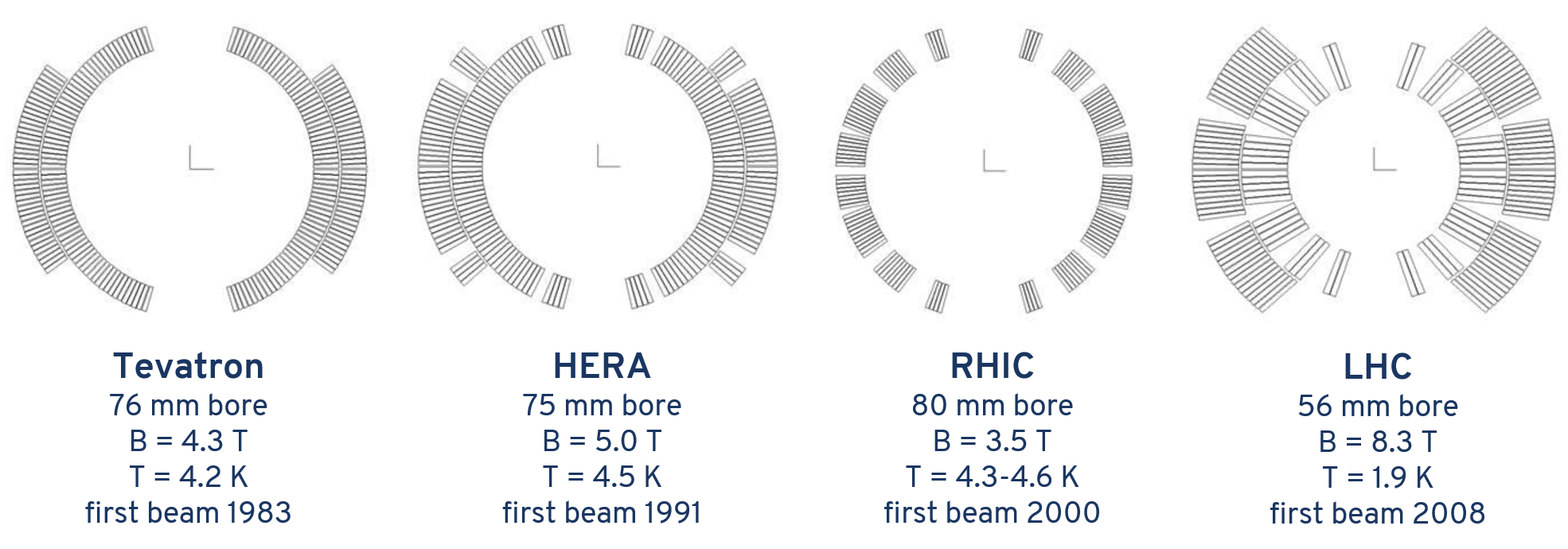}
        \caption{Examples of cos-theta coil distribution in major accelerator dipoles \cite{BotturaMilaneseTommasini2020}.}
	\label{fig:Cos_theta_crosssections}
\end{figure}

The classical coil-dominated layout for accelerator dipoles is the \emph{cos-theta} magnet, in which the ideal current density distribution on the coil circumference varies as $j = j_0 \cos(\theta)$. Tab. \ref{tab:Cos_theta} summarizes ideal current distributions to obtain a dipolar field.
\begin{table}[htbp]
    \centering
    \caption{Comparison of ideal cos-theta current distributions.}
    \label{tab:Cos_theta}
    \begin{tabular}{c m{7.5cm}} 
    \hline\hline
        \raisebox{-0.5\height}{\includegraphics[width=0.3\columnwidth]{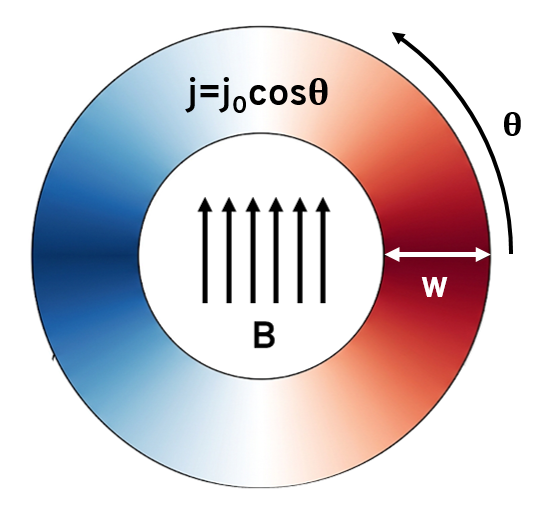}} & 
        Ideal distribution of current density as cos-theta ${B = -\mu_0 j w / 2}$ \\
         & \\ \hline
         \\
        \raisebox{-0.5\height}{\includegraphics[width=0.4\columnwidth]{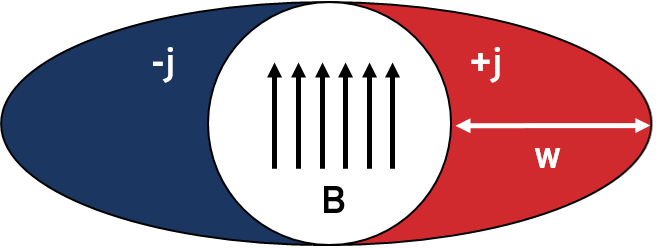}} &
        Ideal uniform current density in the area of two overlapping ellipses ${B = -\mu_0 j w / 2}$ \\
        & \\
        \hline
        \\
        \raisebox{-0.5\height}{\includegraphics[width=0.3\columnwidth]{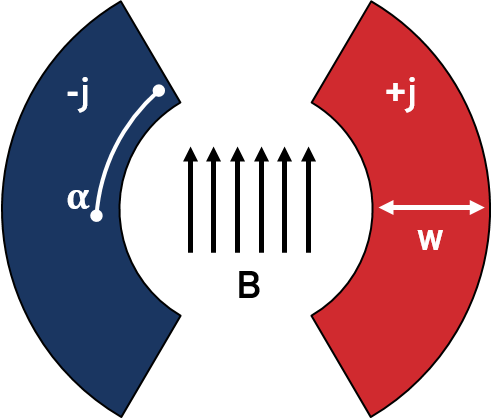}} &
        Ideal uniform current density in 
        sector coils ${B_1 = -\mu_0 (2 j / \pi) w \sin(\alpha)}$ \\
        \hline\hline
    \end{tabular}
\end{table}\\
In practice, these ideal current distributions are approximated by discrete conductor blocks arranged around the aperture, as shown in Fig. \ref{fig:Cos_theta_crosssections}~\cite {BotturaMilaneseTommasini2020}. 
\\

Although these layouts are coil-dominated, the overall dipole cross-sections are not `just coils': substantial mechanical structures, such as collars and yokes, are needed to react the large electromagnetic forces and enclose the magnetic flux.

\subsection{Canted Cos-Theta (CCT) magnets}

\begin{figure}[!htb]
	\centering
	\includegraphics[width=1\columnwidth]{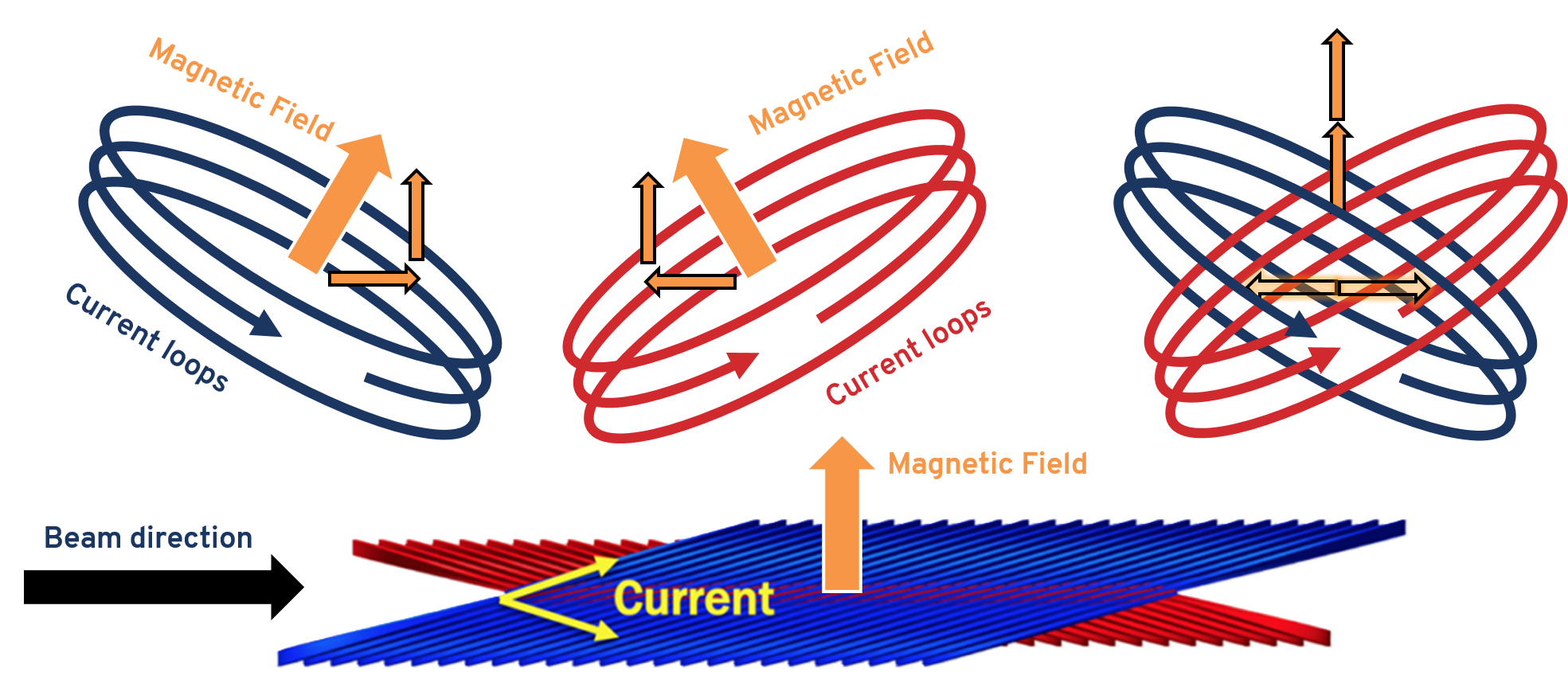}
        \caption{Working principle of the Canted Cos-Theta (CCT) winding: two oppositely tilted current-loop layers combine to generate a net transverse dipole field along the beam direction.}
	\label{fig:CCT}
\end{figure}

An alternative to the classical cos-theta winding is the \emph{Canted Cos-Theta} (CCT) layout, where the cables are arranged as a solenoid, tilted (`canted') with respect to the magnet axis. As illustrated in Fig. \ref{fig:CCT}, superimposing two such opposite windings produces a net dipole field (or any other multipole)  while cancelling the unwanted solenoidal field component~\cite{ShenFajardo2020}. \\

In a CCT magnet, the cable is housed in a mechanical support structure (mandrel) machined to define the conductor path. This construction technique is particularly attractive because, by suitably designing the winding pattern, a CCT magnet can generate essentially any required harmonic, or any superposition of several harmonics (e.g.\ combined dipole and quadrupole fields)~\cite{VanNugteren2017}. Nevertheless, these layouts may be less efficient due to the cancellation of solenoidal components (ampere-turns dedicated to produce not useful field) and may have long heads due to the tilting angle.

\subsection{Toroidal magnets}

Arranging several coils around a common axis generates a toroidal magnetic field, whose azimuthal component follows
\begin{equation}
B_\theta = \mu_0 \frac{N I}{2 \pi R},
\end{equation}

where $R$ is the radial distance from the toroid axis. As we have seen for other magnets, the field of a toroidal magnet can be shaped either by the coils themselves (coil-dominated, typically superconducting), by an iron core (iron-dominated, typically resistive), or by a superferric combination of the two. The space left in between the coils or the iron core can be exploited to let the beam pass through and be deflected by the magnetic field. This feature is at the basis of the toroidal gantry concepts discussed in Section~\ref{sec:gatoroid}.

\section{Superconducting Magnets for Gantries}
\label{sec:gantries}

Gantries are used in particle therapy to deliver the beam to the patient from different angles around a fixed isocentre, where the patient is located. Because the size and weight of a gantry scale with the bending radius, superconducting magnets are considered an enabling technology towards compact gantries, in particular for carbon-ion therapy, where the required beam rigidity is almost three times higher than for protons.
It may be useful to recall that the usual clinical range for protons is between 70-250 MeV (rigidity, $B\rho$ from 1.2 to 2.4 Tm) and for carbon ions is between 140 and 430 MeV/u (rigidity, $B\rho$ from 3.5 to 6.6 Tm). We also recall that the bending radius is the ratio between the beam rigidity and the~applied magnetic field.

\subsection{Superconducting proton gantries}

While superconductivity is widespread in cyclotrons for proton therapy, only two superconducting proton gantries are available on the market. The ProNova \cite{Pronova} gantry uses achromatic superconducting bending sections composed of  4~T dipoles and quadrupoles made of Nb-Ti. The Mevion \cite{Mevion} gantry, by contrast, integrates a gantry-mounted superconducting cyclotron using Nb$_3$Sn conductor operating at 8.5~T, so that the accelerator itself rotates with the gantry rather than being fed by a beam line from a fixed accelerator.

\subsection{Carbon-ion gantries: from resistive to superconducting}

The first clinical carbon-ion gantry, built for the Heidelberg Ion-Beam Therapy centre (HIT, first patient treated in 2012), relies on purely resistive dipoles operating at 1.8~T; owing to the comparatively low field and the high rigidity of carbon ions, the resulting structure is very large, approximately 13~m in diameter and 25~m in length, and with a weight of 600 tons~\cite{Fuchs2008}. Since then, superconducting technology has been adopted to reduce the size and weight of carbon-ion gantries. The QST/Toshiba gantry installed in Chiba (first patient treated in 2018) uses a series of different combined-function Nb-Ti dipole-quadrupole coils operating in the 2.4--2.9~T range~\cite{Iwata2012}. A more recent Toshiba gantry, installed in Yamagata (first patient treated in 2022), uses a similar magnet design operating at 3.5~T, with a length of approximately 8~m and a weight of less than 200~tons~\cite{Takayama2022}.

\subsection{European programmes on superconducting gantry magnets}

Proceeding with the effort of designing compact superconducting ion gantries, three European programmes have been launched to study and prototype different superconducting technologies \cite{Pullia2024,Rossi2024}.\\
CNAO, CERN, INFN, and MedAustron have jointly established the EuroSIG (European Superconducting Ion Gantry) collaboration, aimed at designing and building a novel gantry for hadrontherapy. The~core magnet of the EuroSIG design is a 4~T Nb-Ti curved cos-theta dipole~\cite{Prioli2023}. 
A first winding trial of the demonstrator coil was successfully carried out to validate the winding tooling and procedures~\cite{Prioli2024}, followed by the manufacturing of the full demonstrator magnet, whose curved coil pack and mechanical assembly have been completed~\cite{Prioli2026}.
Within the broader scope of the HITRIplus project \cite{HITRIplusWeb}, a specific work package was dedicated to the development of a 4~T Nb-Ti curved canted-cos-theta dipole ~\cite{HITRIplus2026}. Also in this case, a demonstrator magnet was built to validate the complex design of a curved CCT layout.
As reported in Tab. \ref{tab:euroPrograms}, the EuroSIG and HITRIplus dipoles share the main features, including a 4~T central field, 80~mm of aperture, 1.65~m of radius of curvature, and an angular length of 45$^\circ$ (demonstrators built at 30$^\circ$).\\

A complementary set of straight CCT demonstrators has been developed within the i.FAST project \cite{iFASTweb}. These include a 4~T dipole with a combined-function 5~T/m quadrupole gradient, wound with Nb-Ti rope cables made of six strands, and a pure 4~T dipole wound with a stack of two REBCO tapes~\cite{DeMatteis2026}.

\begin{table}[htbp]
    \centering
    \caption{Comparison of the salient magnet design features for four European programmes on the heavy ion gantry technological development \cite{PrioliIPAC}.}
    \label{tab:euroPrograms}
    \begin{tabular}{|l|c|c|c|c|}
    \hline\hline
        \textbf{Parameter} & \textbf{SIG} & \textbf{HITRIplus} & \textbf{i.FAST LTS} & \textbf{i.FAST HTS} \\
        \hline
        Coil type & Cos--theta & CCT & CCT & CCT \\
        Radius of curvature (m) & 1.65 & 1.65 & $\infty$ & $\infty$ \\
        Dipole field (T) & 4 & 4 & 4 & 4 \\
        Quadrupole gradient (T/m) & 0 & 0 & 5 & 0 \\
        Magnetic length (m) & 0.86 & 0.8 & 0.8 & 0.8 \\
        Aperture (mm) & 80 & 80 & 80 & 80 \\
        Operational temperature (K) & 5 & 4.7 & 4.7 & 20 \\
        Superconductor & Nb-Ti & Nb-Ti & Nb-Ti & REBCO \\
        \hline\hline
    \end{tabular}
\end{table}

\subsection{Alternating Gradient (AG) CCT magnet}

A further evolution of the CCT concept is the \emph{Alternating Gradient} (AG) CCT magnet, which combines a dipole field with alternating focusing and defocusing quadrupole gradients within the same magnet. 
Among others, this combined-function approach was proposed in a collaboration between Lawrence Berkeley National Laboratory (LBNL), the Paul Scherrer Institute (PSI) and Varian Medical Systems to achieve a very large momentum acceptance within a single, comparatively compact magnet in a proton gantry~\cite{Brouwer2017}. 
Such a large momentum acceptance (in the order of $+11\%/-6\%$ in momentum spread) would allow covering all the energy range required for a single treatment section in one or two operational magnetic settings, avoiding the continuous ramping of the magnet and consequent AC losses on the~superconductors.
A reduced-bend prototype of this technology has been produced and tested to validate the winding and field quality of the AG-CCT concept.

\subsection{Superferric magnets: the NIMMS programme}

Within the framework of the CERN Next Ion Medical Machine Study (NIMMS), a superferric option has been studied as an alternative to a fully resistive dipole baseline of the HeLICS synchrotron~\cite{Huttunen2026}.
Considering the same gap size (200 $\times$ 70~mm$^2$), magnetic length (2866~mm), bending angle (60$^\circ$) and bending radius (2737~mm), the superferric design replaces the resistive coil with a REBCO coil. \\

The superconducting coils are designed using CORT (Conductor-on-Round-Tube) cables, obtained by wrapping REBCO tapes around a copper tube \cite{KorchevnyukBottura2025}, where the cryogenic fluid (e.g., liquid nitrogen) circulates.
The main benefit of the superferric option lies in the drastic reduction of resistive power consumption: for the helium-ion design (bore field of 1.65~T), the resistive-coil configuration dissipates about 127~kW, against about 22~kW for the superconducting-coil configuration, i.e.\ a factor of six less; the yoke mass is correspondingly reduced by about 44\%. For the higher-field carbon-ion design (bore field of 2.5~T), the resistive configuration dissipates about 318~kW against about 80~kW for the~superconducting configuration. This reduction in power consumption can be achieved when operating the~REBCO coil at about 65--70~K, using liquid nitrogen, thanks to the temperature margin still available with respect to the critical surface at this operating point.

\subsection{The GaToroid toroidal gantry}
\label{sec:gatoroid}

A conceptually different approach to gantry design is offered by the \emph{GaToroid} concept, originally proposed and patented at CERN by L. Bottura~\cite{Bottura2018}. 
The basic principle consists of the use of the~axially symmetric toroidal magnetic field to bend and focus the particle beam down to the isocentre, providing multiple directions of radiation without the rotation of either the magnet or the patient. The~GaToroid concept is composed of an upstream kicker, which deflects the beam depending on its energy and the~desired direction of treatment, and the toroidal gantry that bends and focuses the beam by using the~free space between the coils. 
Because the toroidal coil system is static, GaToroid replaces the mechanical rotation of a conventional gantry with an electromagnetic selection of the treatment direction, potentially enabling fast, multi-directional treatment without the need for a heavy rotating structure, paving the way toward FLASH treatments~\cite{Bottura2020}. \\

\begin{figure}[!htb]
	\centering
	\includegraphics[width=1\columnwidth]{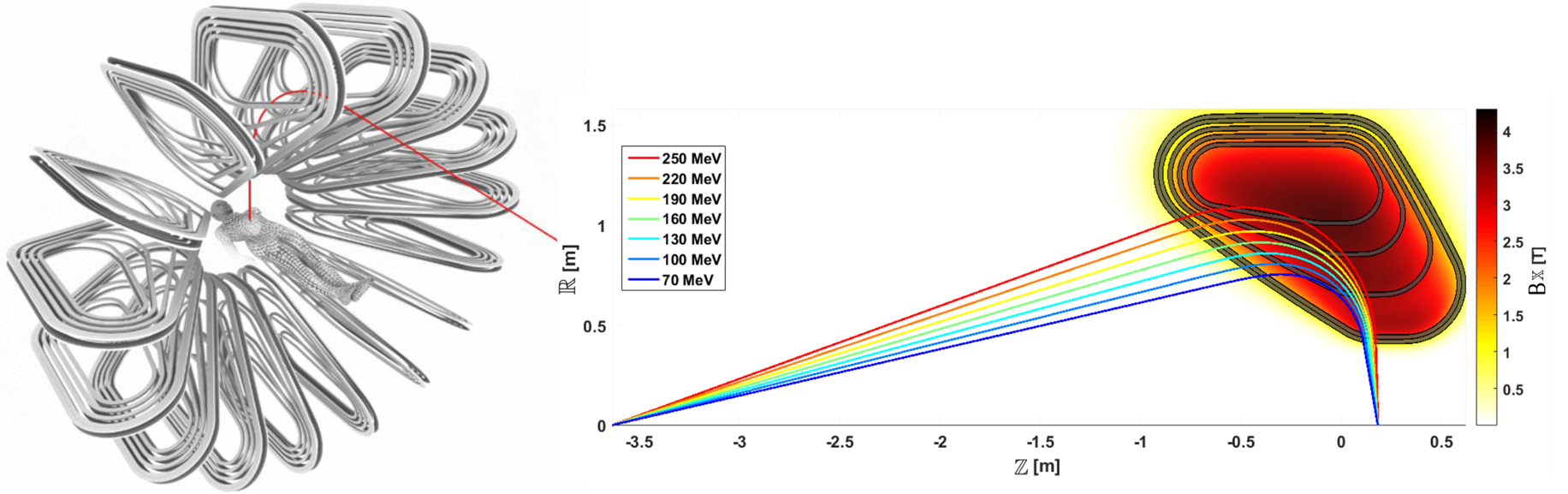}
        \caption{GaToroid torus for protons (left), and proton beam trajectories for energies between 70 and 250~MeV converging at the isocentre (right) ~\cite{Felcini2020}.}
	\label{fig:GaToroid_summary}
\end{figure}

The optimization of the torus geometry and the field profile allows directing the beam down to the tumour location with a very high energy acceptance. 
By adjusting geometrical parameters of the coil and the conductor positions, it was possible to arbitrarily shape the field profile and its dependence on the radius. As a result of the coil optimization shown in Fig. \ref{fig:GaToroid_summary}, it is possible to make the particle trajectories converge on a common point at the isocentre across the full clinical energy range. This concept has been applied to design GaToroid for protons \cite{Felcini2020, Felcini2021Particle} and carbon ions \cite{Felcini2022Ions}. \\

A reduced-scale demonstrator of the GaToroid concept has been designed and built, consisting of a single coil scaled down by a factor of three with respect to the full-size proton design. Two such coils were wound and tested: the first in REBCO tapes and the second in Nb-Ti cables \cite{Felcini2021HTS, Bottura2025}.

While the initial GaToroid design targets protons and ions, a resistive-coil version of the~concept has also been designed for Very High Energy Electron (VHEE) beams, exploiting the reduced beam rigidity of VHEE and therefore building 
significantly smaller, lighter, and less costly gantries~\cite{Bottura2024}. 
\begin{figure}[!htb]
	\centering
	\includegraphics[width=0.8\columnwidth]{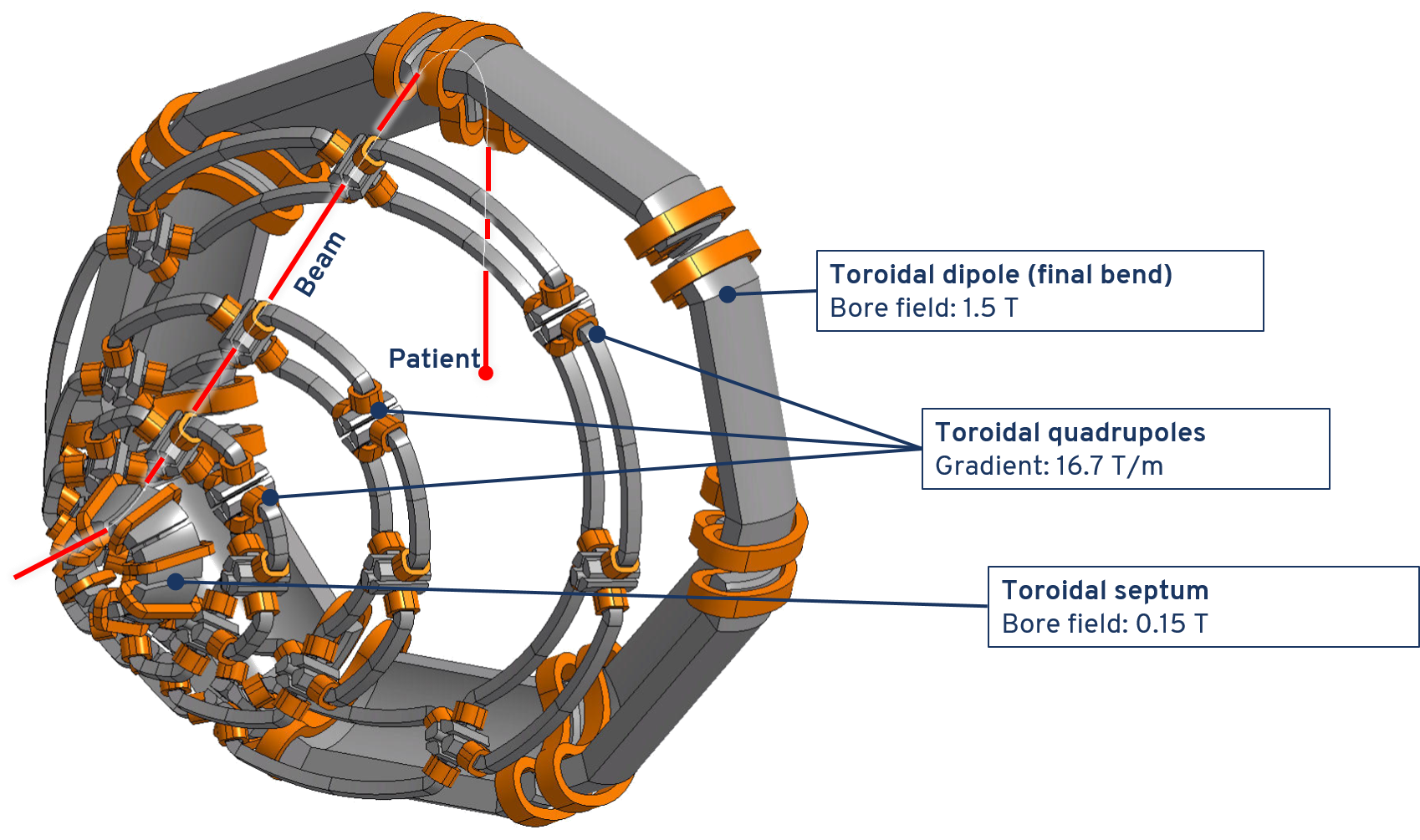}
        \caption{3D model of the VHEE resistive GaToroid.}
	\label{fig:GaToroid_superferric}
\end{figure} \\

The VHEE design shown in Fig. \ref{fig:GaToroid_superferric}, optimized for 7 directions and 200~MeV electron beams, is composed of a toroidal dipole (final bending magnet, bore field of 1.5~T), a set of toroidal quadrupoles (gradient of 16.7~T/m), a toroidal septum magnet (bore field of 0.15~T) and a resonant kicker (similar to one proposed for ELENA \cite{Schoerling2016}), for a total iron weight of about 17~tons and a total coil weight of about 3.7~tons \cite{Bottura2024}. This VHEE design is presently implemented with resistive coils, while investigations are ongoing towards a superferric HTS version aimed at further reducing weight and heat dissipation.

\subsection{Other special applications}

Beyond the magnets and the gantry concepts discussed above, several further applications of superconducting technology are being explored for particle therapy. Among these, we can list: B-dot medical toroidal gantry for protons \cite{Bdot}; fixed-field alternating-gradient (FFGA) gantries with a large momentum acceptance of $\Delta p/p \pm 25\%$, based on an innovative superconducting non-scaling FFAG lattice for carbon-ion therapy~\cite{Trbojevic2011}; Compact PT superconducting scanning magnets operating at 300--1000~T/s and 1--3~T using low-loss Bi-2212 conductor~\cite{Godeke2025}; and static-beam delivery devices for fast scanning proton arc-therapy, in which a combination of bending and guiding coils allows the beam to be steered across an arc of directions without mechanically rotating the treatment head~\cite{Nesteruk2021}.

\section{Concluding Remarks}
\label{sec:conclusions}

This paper illustrates how superconductivity has evolved from an unexpected lab discovery to a mature engineering discipline. What has not changed in this century of developments is the nature of the trade-off: every gain in field, compactness, or weight is bought at the cost of added complexity in conductor, cryogenics, mechanical structure, and quench protection. Therefore, superconducting magnets cannot be designed in isolation, especially for medical applications. They must be considered as one subsystem of a complex engineering ecosystem that includes, above all, the clinical requirements and workflow. \\

The European programmes reviewed in the text turn this trade-off into a concrete design choice. Given nearly identical field, aperture, and magnetic length, the choice between cos-theta and CCT, or between Nb-Ti and REBCO, is really a decision about where to place the complexity: in the coil geometry, the cryogenic margin, or the conductor itself. Superferric magnets and toroidal gantries push the same logic further, showing that significant advancement can be achieved by rethinking the concept of the magnet, or of the whole gantry itself. \\

In conclusion, there is not a unique superconducting technology that is best suited for gantries. Cos-theta, CCT, superferric and toroidal designs are converging, from different directions, on lighter, smaller and more efficient gantries. The choice will depend on both the technological readiness of the~magnets (e.g. LTS vs HTS) and the clinical desiderata of the facility (e.g. ion species and irradiation modalities). Superconductivity, higher achievable fields, and novel magnet topologies are, in this sense, the key enabling technologies for the future of particle therapy.


\end{document}